\documentclass[9pt,technote]{IEEEtran}
\usepackage{comment}
\usepackage{mathtools,amssymb,lipsum}

\usepackage{cuted}
\newcommand\changehighlightone{black}

\usepackage[utf8]{inputenc}
\allowdisplaybreaks
\usepackage[spaces,hyphens]{xurl} 
\usepackage[nocompress]{cite}
\usepackage{amsmath,amssymb,amsfonts}
\usepackage{subcaption}

\usepackage{algorithm}
\usepackage{algpseudocode}
\usepackage{array}
\usepackage{tabularx,booktabs}
\usepackage{graphicx}
\usepackage{caption}
\usepackage{pdfpages}
\usepackage{textcomp}
\usepackage{xcolor}
\usepackage{dsfont}
\usepackage{amsthm}
\usepackage{bm}
\usepackage{array}\usepackage{makecell}

\usepackage{amsmath}
\DeclareMathOperator*{\argmax}{arg\,max}

\usepackage{titlesec}
\titlespacing{\section}{0pt}{9pt}{5pt}
\titlespacing{\subsection}{0pt}{7pt}{3pt}
\titlespacing{\subsubsection}{0pt}{5pt}{2pt}

\usepackage{tikz}
\usepackage{textcomp}
\usepackage{hyperref}
\usepackage{lipsum}

\usepackage{xr}
\makeatletter
\newcommand*{\addFileDependency}[1]{
	\typeout{(#1)}
	\@addtofilelist{#1}
	\IfFileExists{#1}{}{\typeout{No file #1.}}
}
\makeatother

\newcommand\copyrighttext{%
	\footnotesize \textcopyright 2024 IEEE. Personal use of this material is permitted. Permission from IEEE must be obtained for all other uses, in any current or future media, including reprinting/republishing this material for advertising or promotional purposes, creating new collective works, for resale or redistribution to servers or lists, or reuse of any copyrighted component of this work in other works. Digital Object Identifier: https://doi.org/10.1109/TVT.2024.3456114}
\newcommand\copyrightnotice{%
	\begin{tikzpicture}[remember picture,overlay]
		\node[anchor=south,yshift=10pt] at (current page.south) {\fbox{\parbox{\dimexpr\textwidth-\fboxsep-\fboxrule\relax}{\copyrighttext}}};
	\end{tikzpicture}%
}

\begin{document}
	
	\title{Hybrid Network- and User-Centric Scalable Cell-Free Massive MIMO for Fronthaul Signaling Minimization}
	
	\author{Phu~Lai, Wei~Xiang, \IEEEmembership{Senior Member, IEEE}, William~Damario~Lukito, Khoa~Tran~Phan, Peng~Cheng, Chang~Liu, Guoqiang~Mao, \IEEEmembership{Fellow, IEEE}
		\thanks{This article has been accepted for publication by IEEE Transactions on Vehicular Technology. Manuscript received Apr 22, 2024; revised Jul 29, 2024; accepted Sep 04, 2024. Date of publication 0 . 0000; date of current version 0 . 0000.\newline
			(\textit{Corresponding authors: Wei Xiang})}
		\IEEEcompsocitemizethanks{
			\IEEEcompsocthanksitem P. Lai, W. Xiang, W. D. Lukito, K. T. Phan, P. Cheng, and C. Liu are with Cisco-La Trobe Centre for AI and IoT, La Trobe University, Australia. E-mail: \{p.lai, w.xiang, w.lukito, k.phan, p.cheng, c.liu6\}@latrobe.edu.au.
			\IEEEcompsocthanksitem G. Mao is with Xidian Research Institute of Smart Transportation, Xidian University, China. E-mail: g.mao@ieee.org.}
		\thanks{This work was supported in part by the Australian Government through the Australian Research Council's Discovery Projects Funding Scheme under Project DP220101634.}}
	
	\markboth{}%
	{Phu \MakeLowercase{\textit{et al.}}: Hybrid Network- and User-Centric Scalable Cell-Free Massive MIMO for Fronthaul Signaling Minimization}
	\maketitle
	\copyrightnotice
	
	\begin{abstract}
		Cell-free massive multiple-input multiple-output (CFmMIMO) coordinates a great number of distributed access points (APs) with central processing units (CPUs), effectively reducing interference and ensuring uniform service quality for user equipment (UEs). However, its cooperative nature can result in intense fronthaul signaling between CPUs in large-scale networks. To reduce the inter-CPU fronthaul signaling for systems with limited fronthaul capacity, we propose a low-complexity online UE-AP association approach for scalable CFmMIMO that combines network- and user-centric clustering methodologies, relies on local channel information only, and can handle dynamic UE arrivals. Numerical results demonstrate that compared to the state-of-the-art method on fronthaul signaling minimization, our approach can save up to 94\% of the fronthaul signaling load and 83\% of the CPU processing power at the cost of only up to 8.6\% spectral efficiency loss, or no loss in some cases.
	\end{abstract}
	
	
	\section{Introduction}\label{sec:introduction}
	
\begin{figure}
		\centering
		\includegraphics[page=1,scale=0.4]{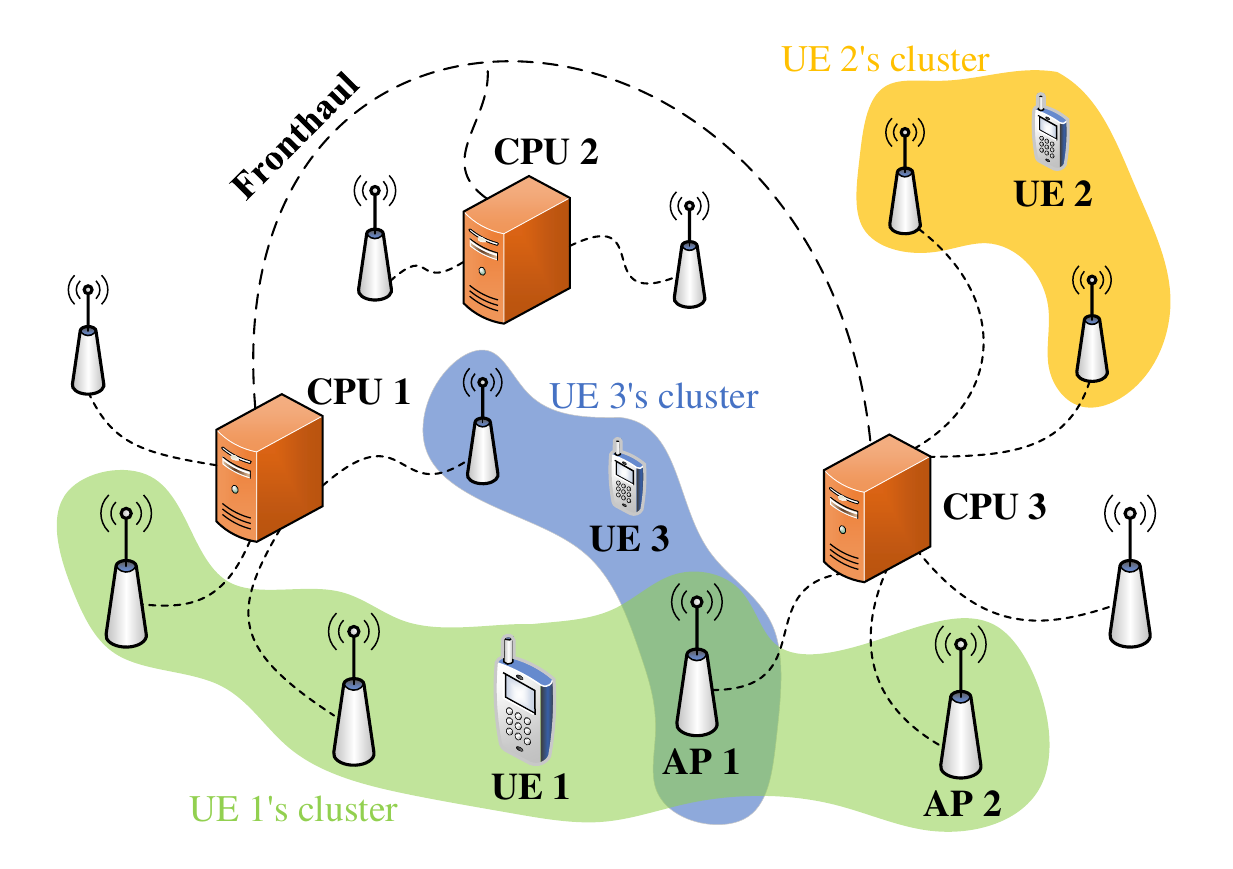}
		\setlength{\belowcaptionskip}{-10pt}
		\caption{A cell-free massive MIMO system.}
		\label{fig:system_model}
	\end{figure}

	\IEEEPARstart{C}{ell-free} massive multiple-input multiple-output (CFmMIMO) is a promising architecture for 6G and beyond. The main novelty of CFmMIMO includes two aspects: the spectral efficiency (SE) analysis with imperfect channel state information \textcolor{\changehighlightone}{\cite{bjornson2020scalable}}, and the integration of the physical layer of mMIMO, ultra-dense network deployment, and coordinated multipoint (CoMP) methods in a scalable manner \textcolor{\changehighlightone}{\cite{ngo2024ultra}}. In CFmMIMO, user equipment (UE) are served by a much larger number of geographically distributed access points (APs), which are coordinated by central processing units (CPUs) through wired or wireless fronthaul (Fig. \ref{fig:system_model}). These APs cooperate phase-coherently to serve users in the same time-frequency resource through time-division duplex (TDD), thereby eliminating the concept of cell boundaries \textcolor{\changehighlightone}{\cite{ngo2017cell}}. Most CFmMIMO research focuses on \textit{user-centric} approaches \textcolor{\changehighlightone}{\cite{buzzi2017cell}}, in which each UE is served by its selected APs (based on some criteria), rather than \textit{network-centric} approaches, which divide APs into separate clusters serving distinct sets of UEs \textcolor{\changehighlightone}{\cite{demir2021foundations}}. Those user-centric approaches that do not take into account inter-CPU coordination often incur very intense inter-CPU signaling, which is not ideal for systems with limited fronthaul capacity. In this correspondence, we propose a hybrid method that combines network- and user-centric clustering and show that its performance is on par with user-centric clustering in terms of SE and fairness (measured by the SE/service uniformity among all UEs) while being superior in terms of scalability with significantly reduced inter-CPU fronthaul signaling and CPU processing power.
	
	Many studies on CFmMIMO do not take system scalability into account and often assume that the fronthaul capacity is unlimited or all APs are connected to a single CPU \cite{ngo2017cell, guenach2020joint, zaher2022learning, kassam2023joint, ammar2021downlink}, which is clearly unscalable. As such, researchers have started investigating the scalability aspects of CFmMIMO \cite{interdonato2019scalability, bjornson2020scalable}, which involves multiple CPUs. However, the fronthaul signaling between CPUs (or network topology and the links between APs and CPUs) is overlooked in those studies. In a practical CFmMIMO system, a UE may be served by APs associated with multiple CPUs, which exchange signals and data to decode the uplink (UL) data and precode the downlink (DL) data. In a large network, this exchange could be very intense and may overload the fronthaul \cite{ranjbar2022cell, ngo2024ultra}. Realizing the limited fronthaul constraint, several studies \cite{freitas2023reducing, li2023joint, ranjbar2022cell} have taken this constraint into consideration. In \cite{li2023joint}, the UE-AP association is pre-determined and the fronthaul is assumed to be flexible, allowing dynamic routing of fronthaul traffic across APs and CPUs through intermediate routing-capable nodes called routers. The authors also consider the finite computing capacity of CPUs. In \cite{ranjbar2022cell}, the authors categorize UEs into local UEs (served by a single CPU, which incurs no inter-CPU signaling) and edge UEs (served by several CPUs, thus inter-CPU signaling is required) based on their geographical distance to the borders of network-centric clusters, which is not robust and ineffective in many scenarios as will be demonstrated in our numerical results. Later on, a cluster refinement method is introduced in \cite{freitas2023reducing} which limits inter-CPU coordination by fine-tuning any arbitrary UE-AP association solutions. We aim to design a new UE-AP association scheme (a.k.a AP selection or UE/AP clustering in the literature) that is more scalable and requires much lower amount of inter-CPU fronthaul signaling and CPU computing power with little to no loss in SE and fairness.
		
	Many existing user association methods are \textit{offline}, or based on network-wide optimization and thus unscalable as the number of UEs grows very large \cite{wei2022user, guo2021joint, vu2020joint}. They (implicitly) require a large amount of fronthaul signaling and a central coordinator, which is aware of all UEs' information in the network, to make a global decision; and their computational complexity grows exponentially with the total number of UEs. It is often unclear how such methods would handle the dynamic situations where there are new UE arrivals. They would either have to re-associate existing UEs or wait for a period of time to associate a new batch of UEs. In contrast, we aim to design a scalable \textit{online} mechanism that associates UEs to APs at the time of their arrivals. This paper focuses on the DL transmission. Our key contributions in this correspondence are:
	\begin{itemize}
		\item A hybrid network- and user-centric clustering scheme for user association that is highly scalable and can handle dynamic UE arrivals. The computational complexity of finding an association decision for a UE is independent of the total number of UEs as the scheme only relies on local channel information.
		\item Numerical results showing that the proposed scheme can save up to 94\% of fronthaul signaling and 83\% of CPU compute power with at most 8.6\% SE loss (or no loss in some cases) compared to the state-of-the-art (SOTA) on fronthaul signaling minimization.
	\end{itemize}
	Our source code (in Python) is accessible at https://github.com/phulai/hybrid-cell-free for reproducibility.
	
	\section{System Model}\label{sec:system_model}
	
	\subsection{Network topology}\label{sec:network_topology}
	\textit{Access points (APs) and central processing units (CPUs):} The system consists of a geographically distributed set $\mathcal{U}$ of $U$ CPUs and set $\mathcal{L}$ of $L$ $N$-antenna APs. Each CPU is connected to a set of APs via ultra-reliable and error-free fronthaul links, which also connect all the CPUs together \cite{demir2021foundations}. There are two types of CFmMIMO implementations categorized by various levels of cooperation between APs/CPUs, namely centralized and distributed implementations \cite{bjornson2020scalable}. This paper adopts the centralized implementation (which has significantly higher SE than the distributed implementation \cite{demir2021foundations}). In the centralized operation, the APs only function as relays that forward their received UL signals to their associated CPUs and forward DL signals generated by CPUs to the UEs. The CPUs are responsible for channel estimation, receive combining, and data detection. We denote by $\mathcal{L}_u \subseteq \mathcal{L}$ the set of APs connected to CPU $u$.
	
	\textit{User equipment (UEs):} The set of $K$ single-antenna UEs is denoted by $\mathcal{K}$. The diagonal matrix decision variable $\mathbf{D}_{kl}$ decides which antennas of AP $l$ may serve UE $k$. In this paper, UE $k$ is served by all antennas of AP $l$ if the AP is assigned to serve that UE, i.e., $ \mathbf{D}_{kl}= \textbf{I}_N$, and $ \mathbf{D}_{kl}= \mathbf{0}_N$ otherwise. We let block-diagonal matrix $\mathbf{D}_k=\text{diag}(\mathbf{D}_{k1},...,\mathbf{D}_{kL})$. We use $\mathcal{L}_k \subseteq \mathcal{L}$ and $ \mathcal{U}_k \subseteq \mathcal{U} $ to denote the sets of APs and CPUs serving UE $k\in\mathcal{K}$. Among $ \mathcal{U}_k $, a CPU is selected as the \textit{master CPU} of UE $ k $ to handle all the signal processing, including channel estimation, receive combining, and data detection. The other CPUs will just act as relays that forward their received signals to the master CPU (on the UL) and to their associated APs serving the UE (on the DL).

	\subsection{Communication model}\label{sec:comms_model}
	The channel $\mathbf{h}_{kl}\in\mathbb{C}^{N}$ between the $k$-th UE and the $l$-th AP is a correlated Rayleigh fading channel, $\mathbf{h}_{kl} \sim \mathcal{N}_{\mathbb{C}}(\mathbf{0},\mathbf{R}_{kl})$, where $\mathbf{R}_{kl}\in\mathbb{C}^{N\times N}$ is the correlation matrix describing the large-scale fading (LSF), which includes spatial channel correlation, pathloss, antenna gains, and shadowing. \textcolor{\changehighlightone}{We considers the absence of dominant line-of-sight (LoS) paths between APs and UEs, which is the case in scenarios such as heavily built-up city centers with rough terrain, many tall buildings and obstacles, hence the use of Rayleigh fading channel model in the system model. For other settings where there is a strong, direct LoS path between the transmitter and receiver (e.g., industrial settings), Rician fading can be used. For a comprehensive performance analysis on Rician fading in CFmMIMO, readers can refer to \cite{zhang2024rate, zhang2024performance}.} The average channel gain between the $k$-UE and an antenna at the $l$-th AP can be expressed by the normalized trace:
	\begin{equation}\label{eq:beta_kl}
		\beta_{kl} = \frac{\text{tr}(\mathbf{R}_{kl})}{N}
	\end{equation}
	The average channel gain $ \beta_{kl} $ is also referred to as LSF coefficient (LSFC) and modeled as \textcolor{\changehighlightone}{\cite{bjornson2017massive, demir2021foundations}}:
	\begin{equation}\label{eq:LSFC}
		\beta_{kl} = \Upsilon - 10 \alpha \log_{10}(d_{kl}) + F_{kl}
	\end{equation}
	where $ \Upsilon $ is the median channel gain at $ 1 $ km reference distance, $ \alpha $ is the path loss exponent, $ d_{kl} $ is the propagation distance between UE $ k $ and AP $ l $, and $ F_{kl} \sim \mathcal{N}(\mathbf{0},\sigma_{\text{SF}}^2) $ is the shadow fading with $ \sigma_{\text{SF}}^2 $ being the shadow fading variance. The collective channel between UE $k$ and all the APs is denoted by $\mathbf{h}_k=[\mathbf{h}_{k1}^{\text{T}}\dots\mathbf{h}_{kL}^{\text{T}}]^{\text{T}}\in\mathbb{C}^{NL}$. The APs' channel vectors are assumed to be independently distributed, thus $\mathbf{h}_{k} \sim \mathcal{N}_{\mathbb{C}}(\mathbf{0},\mathbf{R}_{k})$, with $\mathbf{R}_k=\operatorname{diag}(\mathbf{R}_{k1},\dots,\mathbf{R}_{kL})\in\mathbb{C}^{NL\times NL}$ being the block-diagonal spatial correlation matrix.
	
	The system operates in TDD mode where each $\tau_c$-sample frame (coherence interval) is split into three phases: pilot transmission of $\tau_p$ samples for UL channel estimation (UL training), UL data transmission of $\tau_u$ samples, and DL data transmission of $\tau_d$ samples. In UL training, all the UEs transmit pilot data to the APs to enable channel estimation at the CPUs. There are $\tau_p$ mutually orthogonal $\tau_p$-length pilot signals. \textcolor{\changehighlightone}{We use $t_k \in \{1,...,\tau_p\}$ to denote the index of the pilot allocated to UE $k$\footnote{\textcolor{\changehighlightone}{Pilot assignment is out of the scope of this paper. There are existing pilot assignment methods that are scalable and suitable for our scenario, such as the heuristic proposed in \cite{bjornson2020scalable}, which will be used in our numerical evaluation.}}. The set of UEs sharing the same pilot as UE $k$, including UE $k$, is $\mathcal{K}_{t_k} \subset \mathcal{K}$.} When those UEs transmit the pilot, the signal received at AP $l$ during the entire pilot transmission is:
	\begin{equation}
		\mathbf{y}_{t_k l}^{\text{pilot}}=\sum_{i\in\mathcal{K}_{t_k}}\sqrt{\tau_p p_i}\mathbf{h}_{il}+\textbf{n}_{t_k l}
	\end{equation}
	where $p_i$ is UE $i$'s transmit power, $\textbf{n}_{t_k l}\sim\mathcal{N}_{\mathbb{C}}(\textbf{0}_N,\sigma^2_{\text{UL}} \textbf{I}_N)$ is the additive white Gaussian noise (AWGN). This will then be forwarded to the associated CPUs (and ultimately gathered at one master CPU) via the fronthaul links for channel estimation. Using the received pilot signals, the CPU can calculate all the channel estimate $\{\hat{\mathbf{h}}_{kl}: l\in\mathcal{L}_k\}$. $ \mathbf{h}_{kl} $'s minimum mean squared error (MMSE) estimate, $ \forall k\in\mathcal{K}_{t_k}$, is given by \cite{demir2021foundations}:
	\begin{equation}
		\hat{\mathbf{h}}_{kl}=\sqrt{\tau_p p_k}\mathbf{R}_{kl}\mathbf{\Psi}_{kl}^{-1}\mathbf{y}_{t_k l}^{\text{pilot}}
	\end{equation} 
	with $\mathbf{\Psi}_{kl}=\mathbb{E}\{\mathbf{y}_{t_k l}^{\text{pilot}}(\mathbf{y}_{t_k l}^{\text{pilot}})^\text{H}\} = \sum\limits_{i\in\mathcal{K}_{t_k}}\tau_p p_i \mathbf{R}_{il} + \sigma^2_{\text{UL}} \textbf{I}_N$ being the correlation matrix of $\mathbf{y}_{t_k l}^{\text{pilot}}$.  \textcolor{\changehighlightone}{The mutual interference in $\mathbf{y}_{t_k l}^{\text{pilot}}$ caused by UEs sharing the same pilot ($\mathcal{K}_{t_k}$) is referred to as the pilot contamination. Pilot contamination reduces the channel estimation accuracy and makes the channel estimates $\hat{\mathbf{h}}_{kl}$ of co-pilot UEs $\mathcal{K}_{t_k}$ become correlated, causing coherent interference \cite{bjornson2017massive}. As combining/precoding vectors rely on channel estimates, pilot contamination could reduce the signal-to-interference-plus-noise ratio (SINR) and SE in both UL and DL.}
	
	After the UL training phase, the $i$-th UE transmits intermediary data symbol $s_i\in\mathbb{C}$ to the APs. Let $\mathbf{n}_l\sim\mathcal{N}_{\mathbb{C}}(\mathbf{0}_N,\sigma^2_{\text{UL}} \textbf{I}_N)$ be the noise term, the UL data signal received at the $l$-th AP is:
	\begin{equation}
		\mathbf{y}^{\text{UL}}_l = \sum_{i\in\mathcal{K}}\mathbf{h}_{il} s_i + \mathbf{n}_l
	\end{equation}
	This will be relayed to its associated CPU for centralized data detection. \textcolor{\changehighlightone}{We adopt the partial MMSE (P-MMSE) combining scheme proposed in \cite{bjornson2020scalable}. Compared to the conventional MMSE combining scheme, the calculation of P-MMSE combining vector for a UE $k$ only involves the UEs partially served by the same APs as UE $k$. Its complexity is independent of the total number of UEs $K$, thus it is fully scalable and suitable for large-scale networks, which is the aim of our paper. In terms of performance, the average SE achieved by P-MMSE is within 89\% of the optimal MMSE combining \cite{bjornson2020scalable}.}
	
	In DL data transmission phase, the CPUs compute the normalized precoding vectors $ \mathbf{\bar{w}}_{il}\in\mathbb{C}^N $ for UE $i$ at AP $l$ using the UL channel estimates, generating AP $ l $'s DL signal: 
	\begin{equation}
		\mathbf{x}^{DL}_l = \sum_{i\in\mathcal{K}}\sqrt{\rho_i}\mathbf{D}_{il}\mathbf{\bar{w}}_{il}\varsigma_i 
	\end{equation} 
	where $\rho_i$ is UE $i$'s allocated transmit power and $ \varsigma_i\in\mathbb{C}$ is UE $ i $'s intended data signal ($\mathbb{E}\{\lVert\varsigma_i\rVert^2\}=1$). This DL signal is transmitted to AP $ l $ via fronthaul, which then transmits to its associated UEs. \textcolor{\changehighlightone}{Let $n_k\sim\mathcal{N}_{\mathbb{C}}(0,\sigma^2_{\text{DL}})$ be the receiver noise, the DL signal received by UE $k$ is:
	\begin{equation}
		y^{DL}_k = \sum_{l\in\mathcal{L}} \mathbf{h}^{\text{H}}_{kl} \mathbf{x}^{DL}_l + n_k
	\end{equation} 
	The SINR of UE $k$ in the DL is (Theorems 6.1 in \cite{demir2021foundations}):
	\begin{equation}
		\text{SINR}^{\text{DL}}_k = \frac{\rho_k |\mathbb{E}\{\mathbf{h}^\text{H}_k \mathbf{D}_k \mathbf{\bar{w}}_k\}|^2}{\sum\limits_{i\in\mathcal{K}}\rho_i\mathbb{E}\{|\mathbf{h}^\text{H}_k \mathbf{D}_i \mathbf{\bar{w}}_i|^2\} - \rho_k |\mathbb{E}\{\mathbf{h}^\text{H}_k \mathbf{D}_k \mathbf{\bar{w}}_k\}|^2 + \sigma^2_{\text{DL}}}
	\end{equation}
	where $ \mathbf{w}_k \in \mathbb{C}^{NL}$ is UE $ k $'s precoding vector, which is computed by P-MMSE precoding scheme \cite{bjornson2020scalable} in this paper.} The above communications steps take place after the UE-AP association has been carried out. \textcolor{\changehighlightone}{We use this well-known communication model as representative, our UE-AP association approach can also be applied to other communication models as it only requires the information on $\mathbf{R}_{kl}$}.
	
	\subsection{Problem formulation}\label{sec:prob_formulation}
	\textcolor{\changehighlightone}{Our objective is to assign UEs to APs such that 1) the fronthaul signaling load is minimized, 2) while maximizing the sum SE. The fronthaul signaling load, quantified by the number of complex scalars transmitted between CPUs, is measured as follows.} In each coherence block, in the UL, the $l$-th AP transmits $ \tau_p N $ and $ \tau_u N $ complex scalars representing pilot signal $\mathbf{y}_{t_k l}^{\text{pilot}}$ and received data signal $ \mathbf{y}^{\text{UL}}_l $, respectively, to its associated CPU. In the DL, the CPU returns $ \tau_d N $ complex scalars to AP $l$ representing DL signal $ \mathbf{x}^{DL}_l $ \cite{demir2021foundations, bjornson2020scalable}. This CPU will then need to relay some of its received signals (complex scalars) to other CPUs, which incurs inter-CPU fronthaul signaling. Take Fig. \ref{fig:system_model} for instance, UE 1 is served by APs linked to CPU 1 and CPU 3 with CPU 1 being the master CPU, which handles all the signal processing. CPU 3 acts as a relay that forwards the signals received by AP 1 and AP 2 to CPU 1 via fronthaul. UE 3 is served by APs also linked to CPU 1 and CPU 3, with CPU 1 also being the master CPU. As the signals previously relayed to CPU 1 by CPU 3 (for UE 1) already include UE 3's signals, no extra inter-CPU signaling is required for UE 3. UE 2 is served by APs linked to one single CPU and does not cause any inter-CPU fronthaul signaling. \textcolor{\changehighlightone}{Formally, the fronthaul load minimization objective can be expressed as follows:
	\begin{subequations}
		\begin{align}
			\underset{\{\mathbf{D}_k:\forall k\}}{\text{minimize}} & \sum_{u\in\mathcal{U}} \sum_{l\in\mathcal{L}^{\dagger}_u} N \tau_c \label{eq:obj_fronthaul}\\
			\text{subject to } 
			& \mathcal{K}_u=\{k\in\mathcal{K}:\text{tr}(\mathbf{D}_{kl})\ge 1, l\in\mathcal{L}_u\}, \forall u\in\mathcal{U} \label{eq:cnstr_UEs_u}\\
			& \mathcal{K}^{\dagger}_u = \{k\in\mathcal{K}_u: \argmax_{i\in\mathcal{U}_k}|\mathcal{L}_k\cap\mathcal{L}_i|=u \}, \forall u\in\mathcal{U} \label{eq:cnstr_UEs_master_u}\\
			& \mathcal{L}^{\dagger}_u = \{l:l\in\mathcal{L}_k\setminus\mathcal{L}_u, k\in\mathcal{K}^{\dagger}_u \} \label{eq:cnstr_APs_UEs_master_u}, \forall u\in\mathcal{U}
		\end{align}
	\end{subequations}
	Objective \eqref{eq:obj_fronthaul} minimizes the inter-CPU fronthaul signaling load, measured by the number transmitted complex scalars. Set $ \mathcal{K}_u$ \eqref{eq:cnstr_UEs_u} denotes the set of UEs served by APs associated with CPU $u$. Set $\mathcal{K}^{\dagger}_u$ \eqref{eq:cnstr_UEs_master_u} denotes the set of UEs whose master CPU is CPU $u$. The master CPU of a UE is defined as the CPU that has the most number of associated APs serving the UE. Set $\mathcal{L}^{\dagger}_u$ \eqref{eq:cnstr_APs_UEs_master_u} denotes the set of APs serving UEs whose master CPU is CPU $u$, excluding APs associated with master CPU $u$. Note that $\mathcal{L}^{\dagger}_u$ may contain APs associated with multiple CPUs. APs associated with the master CPU are excluded from this set because the signals received by those APs do not have to be relayed to any other CPU.}
	
	\textcolor{\changehighlightone}{Following the primary objective of minimizing the fronthaul signaling load, we aim to maximize the sum-SE of all UEs. 
	\begin{subequations}
		\begin{align}
			\underset{\{\mathbf{D}_k:\forall k\}}{\text{maximize}} & \sum_{k\in\mathcal{K}} \frac{\tau_d}{\tau_c} \log_2\big(1+\text{SINR}^{\text{DL}}_k\big) \label{eq:obj_SE}
		\end{align}
	\end{subequations}}

	\section{Proposed Hybrid Network- and User-centric Online User Association} \label{sec:proposed_solution}
	Our proposed approach is summarized in Algorithm \ref{alg:algo_ours}. When UE $k$ joins the network, it measures its channel quality to every network-centric cluster, which is a set of APs associated with the same CPU (Line 2). The channel quality $\beta_{ku}$ between UE $k$ and CPU $u$'s cluster is approximated by the sum of large-scale fading coefficients (LSFC) $\beta_{kl}$ between the UE and all APs in the cluster: $\beta_{ku} = \sum_{l\in\mathcal{L}_u}\beta_{kl}$.
	
	The next step is the key part of our solution - identify if this UE would be likely to receive sufficient SE from the APs linked to the same CPU, or if the UE would be better off being served by APs linked to multiple CPUs, which requires inter-CPU fronthaul signaling. To do so, we calculate z-scores $z_{ku}$ for each $\beta_{ku}, \forall u\in\mathcal{U}$ (Line 3): \textcolor{\changehighlightone}{ $z_{ku} = \frac{\beta_{ku} - \mu_k^{\text{LSFC}}}{\sigma_k^{\text{LSFC}}}$, where $\mu_k^{\text{LSFC}} = \frac{\sum_{u\in\mathcal{U}}\beta_{ku}}{U}$ and $\sigma_k^{\text{LSFC}}=\sqrt{\frac{\sum_{u\in\mathcal{U}}(\beta_{ku} - \mu_k^{\text{LSFC}})^2}{U}}$.}
	
	Z-score is a statistical measure measuring how far a value is from the mean of a group of values; e.g., a z-score of 1 means that $\beta_{ku}$ is one standard deviation away from the mean of all $\beta_{ku}$\footnote{In a very strict statistical sense, calculating z-scores makes more sense for normally distributed data. While we cannot guarantee the distribution of $\beta_{ku}$ being a normal distribution, our numerical results demonstrate that this method is quite effective.}. We set a threshold $\epsilon$ for the z-scores. If there is only one $\beta_{ku}$ exceeding this threshold and also being the highest $\beta_{ku}$\footnote{Checking if $\beta_{ku}$ is the highest is required because exceeding this threshold could also mean that $\beta_{ku}$ is $\epsilon$-standard deviation \textit{below} the mean, which infers poor channel quality.} (Line 4), then this cluster $u$ stands out from all the other network-centric clusters, i.e., channel quality well above the others, and this UE is likely to be sitting in the middle of this network-centric cluster $u$, and likely to receive good service from APs belonging to just this cluster (\textit{network-centric clustering}). In this case, we adopt the technique proposed in \cite{ngo2017total} to determine the selective best APs, which collectively contribute at least $\delta$\% of the total LSFC of all APs connected to this cluster $u$, to serve UE $k$.
	
	\begin{algorithm}
		\caption{Hybrid online user association (HybridUA)} \label{alg:algo_ours}
		\begin{algorithmic}[1]
			\For{each UE $ k \in \mathcal{K}$}
			\State Measure channel quality $\beta_{ku}, \forall u\in\mathcal{U}$
			\State Calculate z-scores $ z_{ku}, \forall u\in\mathcal{U}$
			\If{$ |\{z_{ku}|z_{ku}\geq \epsilon, \forall u\in\mathcal{U}\}|=1 $ and $ u=\argmax_u \beta_{ku} $}
			\State $ \mathcal{U}_k^{\text{top}} = \argmax_u \beta_{ku} $
			\Else
			\State $ \mathcal{U}_k^{\text{top}} $ is top $ \upsilon $ CPUs with highest $ \beta_{ku} $ 
			\EndIf
			\State Find APs $ \mathcal{L}_k $ such that $ \frac{\sum_{l\in\mathcal{L}_k}\beta_{kl}}{\sum_{u\in\mathcal{U}_k^{\text{top}}}\sum_{l\in\mathcal{L}_u}\beta_{kl}}\geq \delta\% $
			\State $ \mathbf{D}_{kl}= \textbf{I}_N, \forall l\in\mathcal{L}_k $
			\EndFor
		\end{algorithmic}
	\end{algorithm}
	
	In other cases where multiple $\beta_{ku}$ exceed $\epsilon$, does it means that those clusters have equally good channel quality? Not necessarily. It just means that this UE is possibly sitting in the middle/intersection of those clusters. The same applies to the cases where no $\beta_{ku}$ exceeds $\epsilon$. In such cases, it would more beneficial for the UE to receive service from APs linked to different CPUs/clusters (\textit{user-centric clustering}). We define a parameter $\upsilon$ to enforce the maximum number of CPUs/clusters a UE can be assigned to. Again, we apply the technique proposed in \cite{ngo2017total} to select APs that collectively contribute at least $\delta$\% of the total LSFC of all APs connected to the top-$\upsilon$ clusters, which are the clusters with highest sum LSFC, to serve UE $k$. Among those $\upsilon$ CPUs, one with more APs selected to serve the UE will be assigned as the master CPU. The other CPU(s) simply just relay their received UE signals to this master CPU for signal processing. The time complexity of finding an association decision for each UE (Lines 2-10 of Algorithm \ref{alg:algo_ours}) is $ \mathcal{O}(U\log L) $, which is independent of the total number of UEs $ K $. Its complexity does not grow with the number UEs in the network and thus is scalable. 
	
	\textcolor{\changehighlightone}{Algorithm \ref{alg:algo_ours} involves three thresholds, $\epsilon$, $\upsilon$, and $\delta$, controlling the trade-off between SE and fronthaul signaling load. Decreasing $\epsilon$, increasing $\upsilon$ and $\delta$ will increase the SE at the cost of extra fronthaul signaling. However, our simulations show that this SE improvement is often very marginal (up to 2\%) and not worth the extra fronthaul signaling load (up to 70\%). It appears there is a sweet spot for these thresholds (LSFC contribution threshold $\delta=95$\% and per-UE CPU/cluster limit $ \upsilon=2 $ CPUs/clusters) that is robust and consistent across all simulation settings (Sect. \ref{sec:numerical_results}).}
	
	\textcolor{\changehighlightone}{In summary, the aspects of fronthaul load minimization and SE maximization are interconnected throughout the proposed algorithm via channel quality (LSFC). First, each UE identifies if there is any network-centric cluster whose sum LSFC is significantly higher than that of all other clusters (Lines 2-5 of Algorithm 1). In this case, inter-CPU fronthaul signaling is not required while the UE is still likely to achieve high SE. If there is no such cluster, the UE will opt to be served by APs linked to top-$\upsilon$ clusters with the strongest channels in order to achieve better SE.}
	
	\section{Numerical Results} \label{sec:numerical_results}
	\subsection{Simulation settings and benchmark approaches}\label{sec:sim_settings}
	Our CFmMIMO network contains $K$ UEs, $L$ $N$-antenna APs, and $U$ CPUs. The AP-CPU association is generated via k-means clustering based on geographical locations. The UEs and APs are uniformly randomly distributed in a square area. The median channel gain at $ 1 $ km reference distance $ \Upsilon=-148.1 $ dB, path loss exponent $ \alpha=3.76 $, shadow fading standard deviation $ \sigma_{\text{sf}}=10 $. The bandwidth is $ 20 $ MHz with receiver noise power $ \sigma^2 = -94$ dBm consisting of thermal noise and $ 7  $ dB noise figure. We set $\tau_p=10$ and $\tau_d=190$ when evaluating DL. All UEs have equal transmit power $p_k=100$mW in the UL for all approaches in the benchmark. In the DL, we adopt a scalable centralized fractional power allocation (\cite{demir2021foundations}, Sec. 7.2.2). For pilot allocation, all approaches apply the scalable heuristic proposed in \cite{bjornson2020scalable}. In all simulations, z-score threshold $ \epsilon $ is set to 0.4, the number of antennas per AP (N) is 4, the total LSFC contribution threshold $\delta$ is 95\%, \textcolor{\changehighlightone}{and per-UE CPU limit $\upsilon$ is 2 CPUs.}
	
	We evaluate our proposed approach (HybridUA) against the following user association schemes. Note that we only consider schemes that are designed to be scalable.
	\begin{itemize}
		\item SCF1: Scalable cell-free (SCF) scheme proposed in \cite{bjornson2020scalable}, based on dynamic cooperation clustering framework \cite{bjornson2011optimality} and does not take into account the cooperation between CPUs.
		\item SCF2: Scalable cell-free scheme proposed in \cite{interdonato2019scalability}. Each UE is served by all APs associated with a fixed number of CPUs, which is 2 in our simulations. 
		\item SCF1lim: The SOTA on inter-CPU coordination minimization, proposed in  \cite{freitas2023reducing}. It fine-tunes any arbitrary UE-AP association solutions, which are produced by SCF1 in this paper, by dropping the UEs coordinated by multiple CPUs and have the worst channels.
		\item Border: It limits the inter-CPU coordination based on UEs' distance to the borders (100m in this paper) of network-centric clusters, proposed in \cite{ranjbar2022cell}. UEs near the border are served by multiple CPUs and require inter-CPU signaling while the other UEs do not.
		\item LLSFB (Largest-Large-Scale-Fading-Based): A network-centric variant of the LLSFB user association scheme proposed in \cite{ngo2017total} that we came up with. For each UE, we first identify which network-centric cluster has the greatest sum LSFC, then select APs that collectively contribute at least $\delta$\% of the total LSFC of all APs in this cluster to serve the UE.
		\item Nearest: This naive network-centric scheme assigns each UE to all APs associated with the UE's nearest network-centric cluster. Nearest and LLSFB do not incur inter-CPU fronthaul signaling.
	\end{itemize}
	Beside the key performance metrics, which are inter-CPU fronthaul signaling load and per-UE SE, we also report the numbers of APs serving a UE, the numbers of UEs served by an AP or a CPU, and fairness. The fairness indicates whether the UEs are receiving equal SE. To quantify the fairness, we utilize Jain's fairness index, calculated as $(\sum_{k\in\mathcal{K}}\text{SE}_k)^2 / (K \sum_{k\in\mathcal{K}}\text{SE}_k^2)$, which ranges from 0 to 1, with 1 being the fairest. UL results, which are fairly similar to DL results due to channel reciprocity, are omitted for brevity. 
	\begin{figure}
		\centering
		\includegraphics[page=1,scale=0.4]{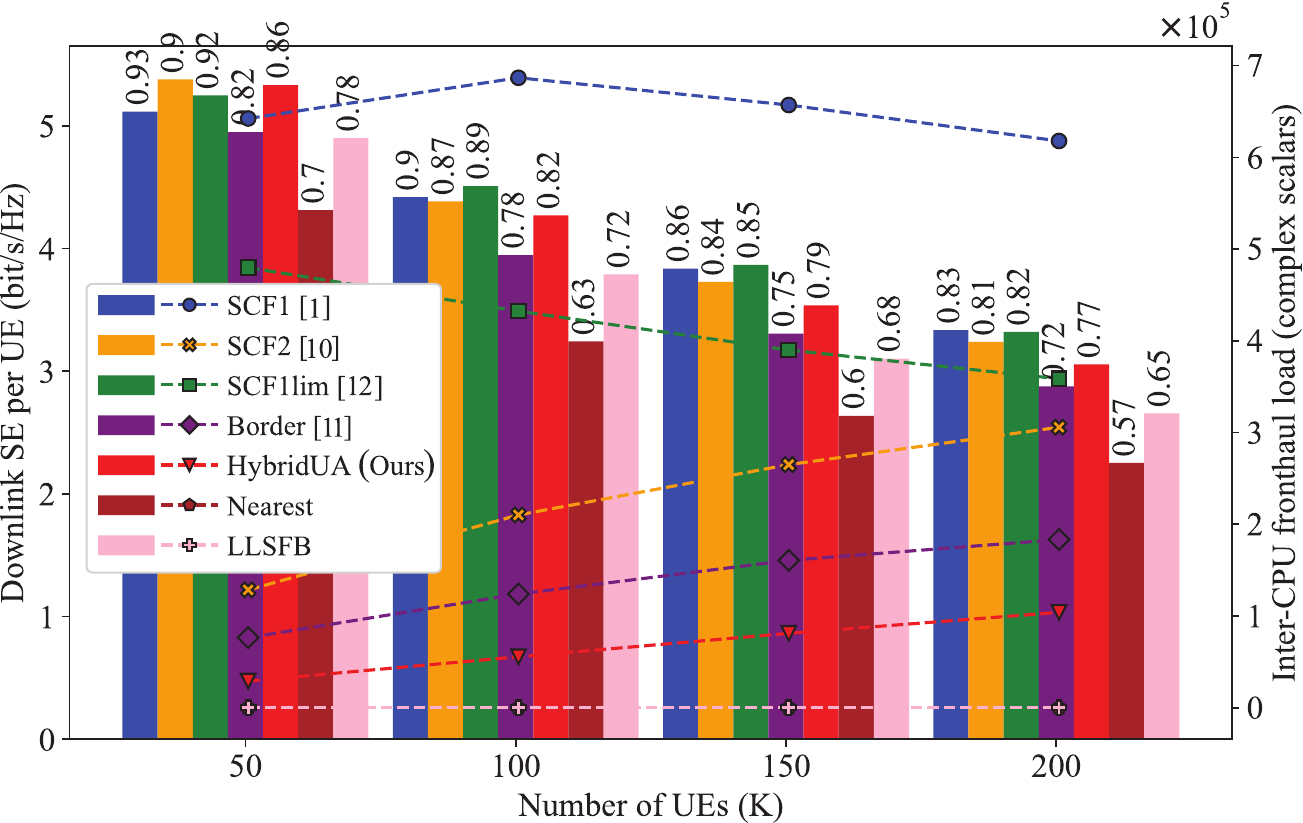}
		\caption{\textcolor{\changehighlightone}{The bars correspond to the left y-axis, which represents per-UE DL SE. The lines correspond to the right y-axis, which represents inter-CPU fronthaul signaling load. The number above each bar is the Jain's fairness index of the UEs' DL SE.}}
		\label{fig:varying-K-SE}
	\end{figure} 
	
	\begin{figure}
		\centering
		\includegraphics[page=1,scale=0.4]{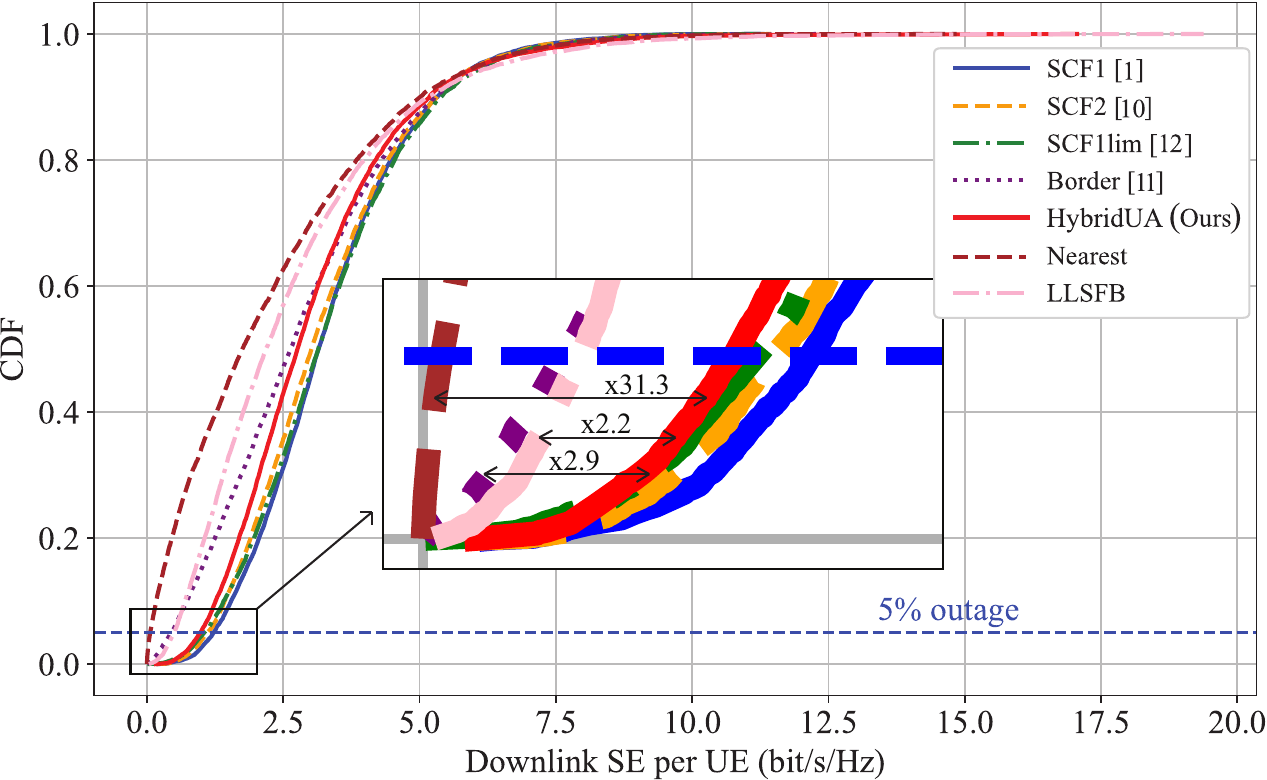}
		\caption{CDF of per-UE DL SE.}
		\label{fig:varying-K-CDF}
	\end{figure} 
	\subsection{Impact of network traffic (number of UEs)}\label{sec:sim_K} 
	This simulation consists of 200 APs and 40 CPUs over a 8$\text{km}^2$ area (100 antennas/$\text{km}^2$, 5 APs/CPU). The number of UEs is varied between 50 and 200 (6.25-25 UEs/$\text{km}^2$). As shown in Fig. \ref{fig:varying-K-SE}, compared to SCF1lim (the latest work on inter-CPU signaling minimization), at the cost of a small loss in per-UE DL SE (up to 8.6\% loss), HybridUA can save 71-94\% fronthaul signaling. In some cases ($K=50$), we even observe no loss in SE. As the number of UEs increases, the fronthaul load of HybridUA also increases at a relatively slow rate. Interestingly, the fronthaul load of SCF1 and SCF1lim decreases, which happens because those schemes explicitly limit the number of UEs per AP; the more UEs, the fewer APs (and CPUs) serving each UE. Nevertheless, the fronthaul load of SCF1 and SCF1lim are still significantly higher than HybridUA. SCF2 does help with inter-CPU fronthaul load as each UE is served by a fixed number of network-centric clusters. However, its fronthaul load is still three times higher than HybridUA while only offering a very marginal SE improvement. LLSFB and Nearest, although incur no inter-CPU signaling, losing 9-15\%, 24-36\% SE compared to HybridUA, respectively. Border achieves 6-8\% lower SE than HybridUA while uses 43-62\% higher fronthaul signaling.
	
	Jain's fairness index (the number above each bar in Fig. \ref{fig:varying-K-SE}) provides a quantitative measurement of the fairness/uniformity of the UEs' SE. HybridUA's fairness is at most 0.08 lower than the fairest scheme, and 0.06 lower than SCF1lim. Nearest's and LLSFB's fairness are up to 0.2 and 0.12 lower than HybridUA, respectively. This shows that HybridUA can provide a uniform service to the UEs. 
	
	Fig. \ref{fig:varying-K-CDF} illustrates the cumulative distribution function (CDF) of per-UE DL SE with 200 UEs, which confirms HybridUA's service uniformity. Compared to HybridUA, LLSFB, Border, and Nearest have $ 2.2\times $, $ 2.9\times $, and $ 31.3\times $ lower per-UE SE in the 5\% outage (top 5\% most unfortunate UEs), respectively. This shows how HybridUA can make a substantial improvement to the most unfortunate UEs.
		
\begin{figure}
		\centering
		\includegraphics[page=1,scale=0.4]{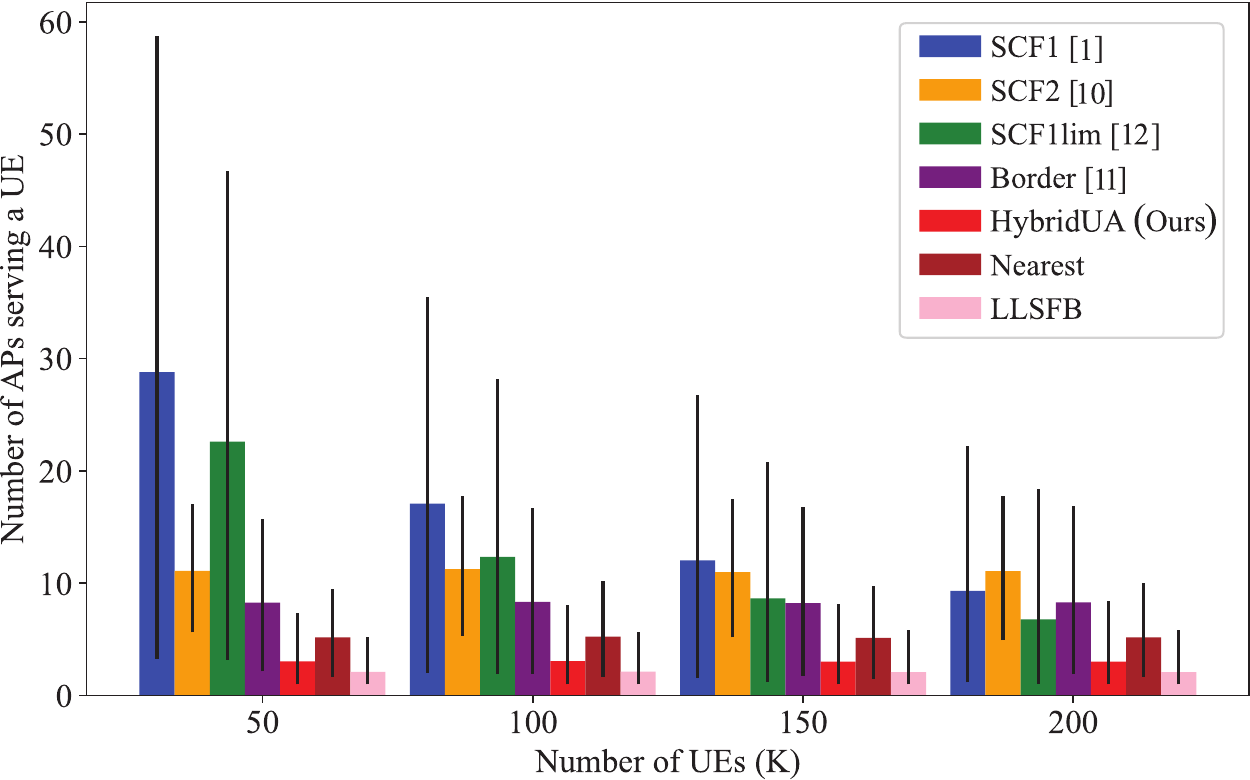}
		\caption{Average number of APs serving a UE (the colored bars). The error bars illustrate the lowest/highest numbers.}
		\label{fig:varying-K-AP-per-UE}
	\end{figure}

\begin{figure}
		\centering
		\includegraphics[page=1,scale=0.4]{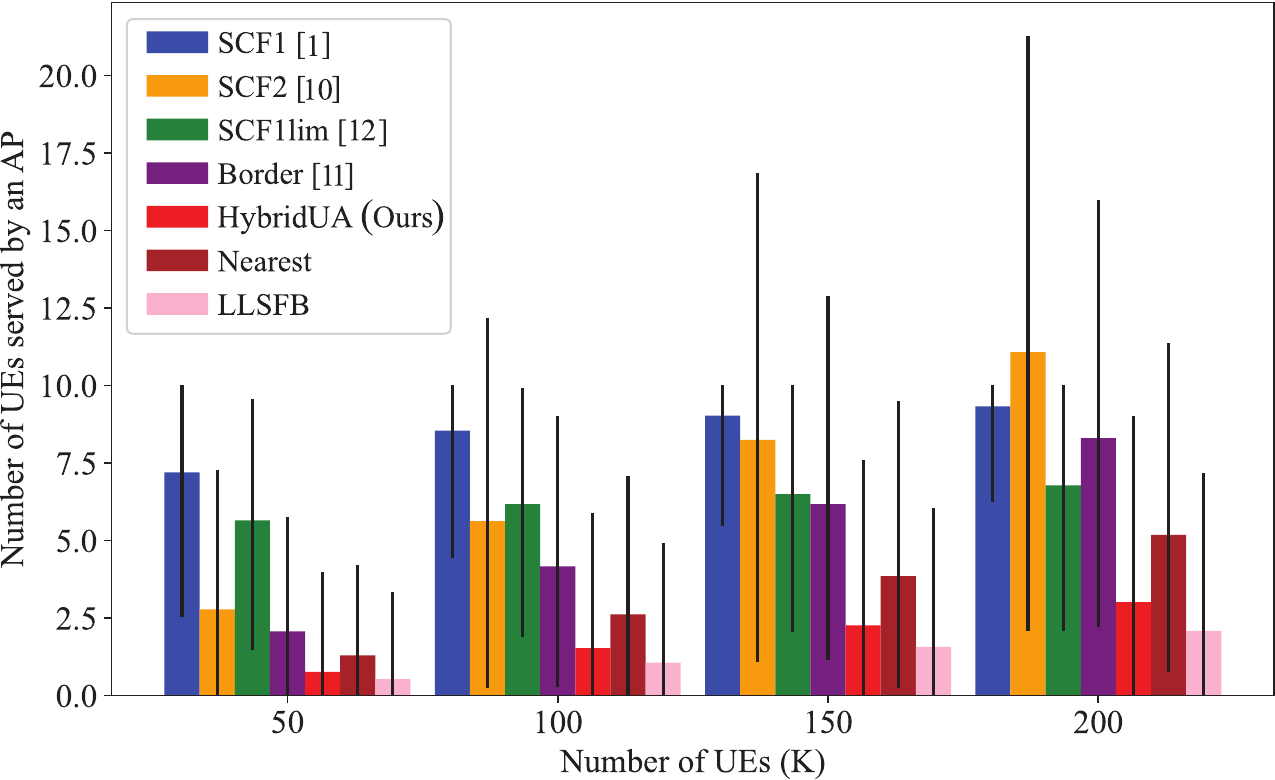}
		\caption{Average number of UEs served by an AP (the colored bars). The error bars illustrate the lowest/highest numbers.}
		\label{fig:varying-K-UE-per-AP}
	\end{figure}

\begin{figure}
		\centering
		\includegraphics[page=1,scale=0.4]{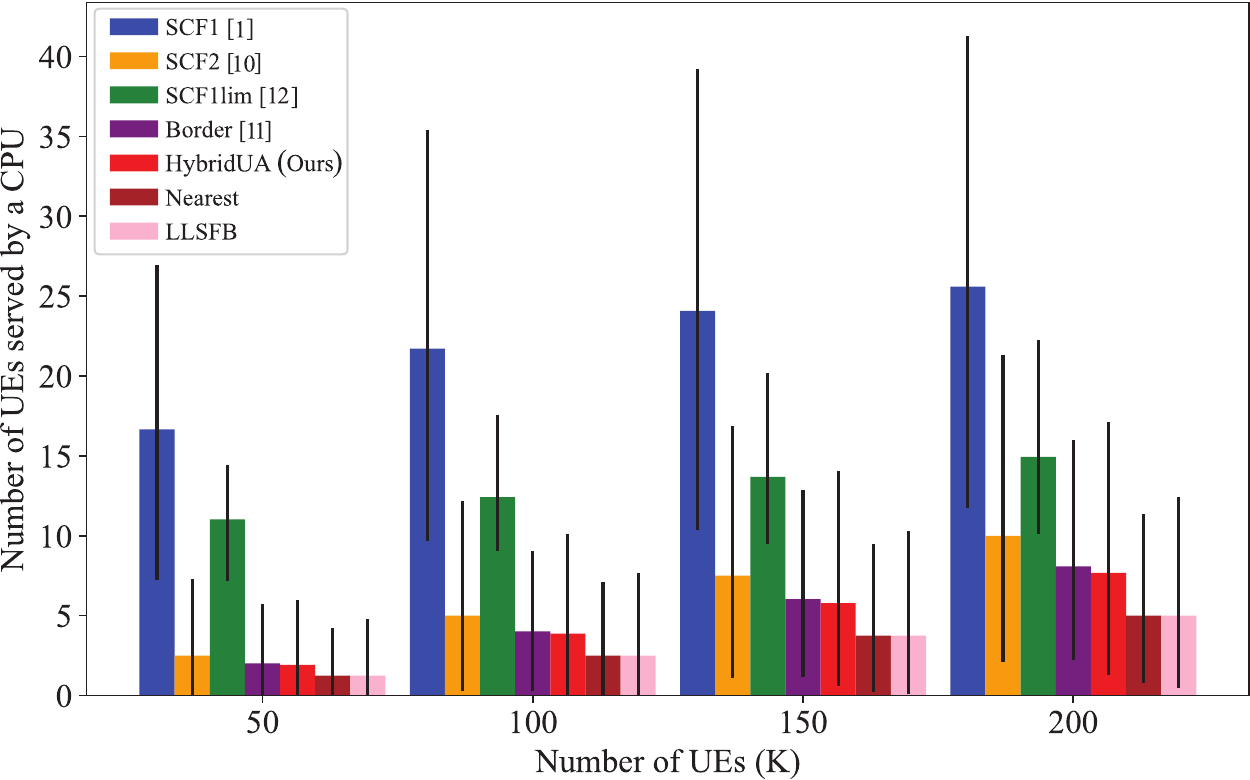}
		\caption{Average number of UEs served by a CPU (the colored bars). The error bars illustrate the lowest/highest numbers.}
		\label{fig:varying-K-UE-per-CPU}
	\end{figure}

	Fig. \ref{fig:varying-K-AP-per-UE} depicts that we do not need a very large number of APs to serve each UE in order to achieve satisfactory SE. In fact, on average, HybridUA requires only 3 APs (8 APs at most) per UE regardless of the number of UEs being simulated - significantly fewer than all other schemes except LLSFB. This is the reason why our scheme performs so well. The other schemes may require up to $ 10\times $ more APs per UE on average only to achieve a 8.6\% increase in SE. Fig. \ref{fig:varying-K-UE-per-AP} shows that while HybridUA does not explicitly limit the number of UEs per AP like SCF1 and SCF1lim do, the number of UEs per AP by HybridUA is much lower than that of SCF1 and SCF1lim. \textcolor{\changehighlightone}{Thus, the effect of pilot contamination is minimized as the UEs are less likely to share a pilot.} This also proves the scalability of our approach with regards to processing power of APs. Fig. \ref{fig:varying-K-UE-per-CPU} shows that on average, HybridUA requires each CPU to serve just a few UEs - 49-83\% fewer UEs per CPU than SCF1lim, saving a substantial amount of computing power on CPUs (for channel estimation, receive combining, and data detection) and making HybridUA even more scalable than SCF1 and SCF1lim.
	
	\subsection{Impact of the number of CPUs and area size}\label{sec:sim_ap_cpu} 
\begin{figure}
		\centering
		\includegraphics[page=1,scale=0.38]{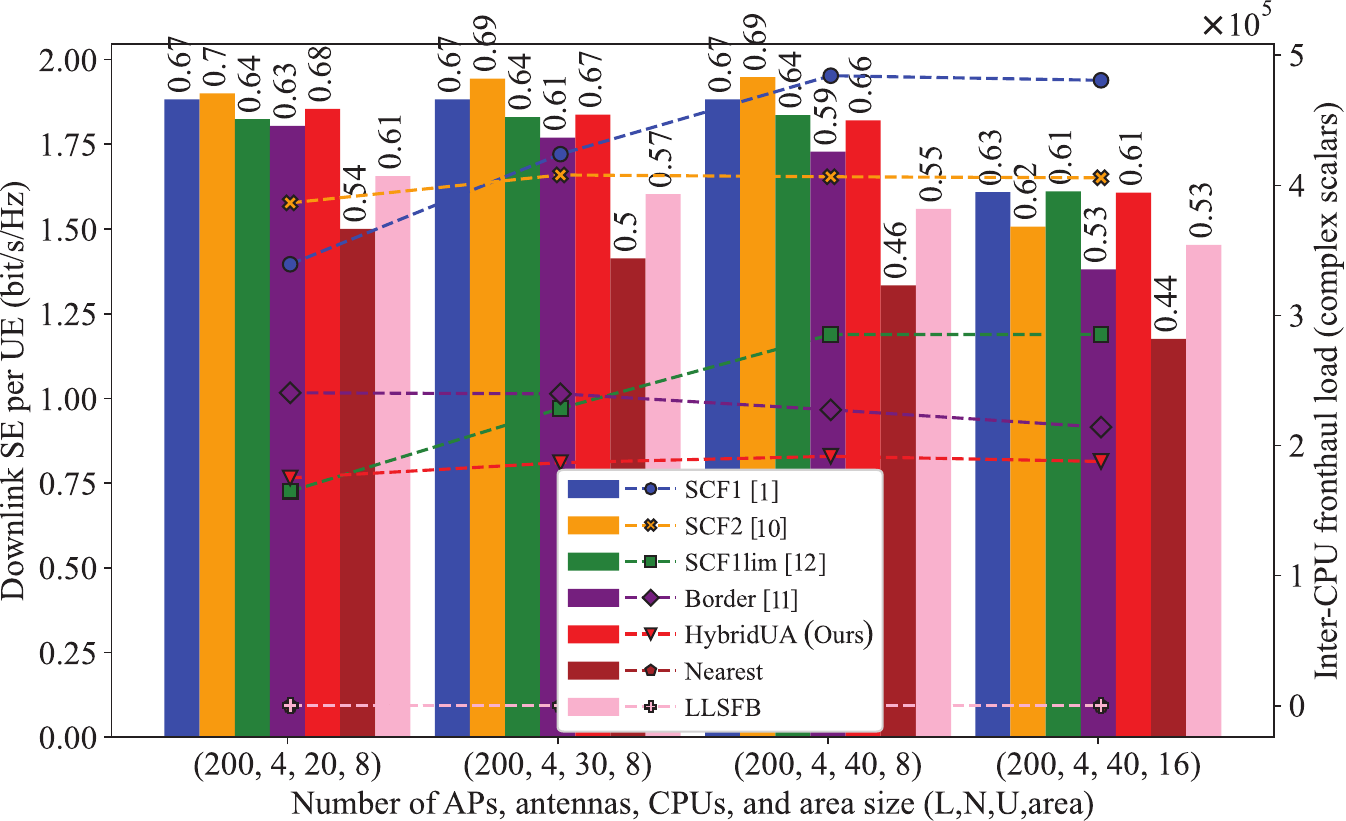}
		\caption{\textcolor{\changehighlightone}{The bars correspond to the left y-axis, which represents per-UE DL SE. The lines correspond to the right y-axis, which represents inter-CPU fronthaul signaling load. Each four-number tuple on the x-axis represents (total number of APs, number of antennas per AP, total number of CPUs, area size in $\text{km}^2$). There are 200 APs, 400 UEs, and 20/30/40 CPUs in a 8$\text{km}^2$ area (first three tuples), and 40 CPUs in a 16$\text{km}^2$ area (last tuple). The number above each bar is the Jain's fairness index of the UEs' DL SE.}}
		\label{fig:varying-CPU-SE}
	\end{figure}

	The first three groups of bars in Fig. \ref{fig:varying-CPU-SE} shows the results of various AP/CPU ratios, ranging from 10 to 5 APs per CPU. Compared to the SOTA (SCF1lim), HybridUA only loses up to 1.7\% SE while saves up to 33\% fronthaul load. HybridUA's fronthaul load remains virtually unchanged as the number of CPUs increases; whereas the load incurred by SCF1 and SCF1lim grows steadily. SCF2 and Border's fronthaul load also remains relatively unchanged. Nevertheless, SCF2's fronthaul load is still twice as high as HybridUA's load only to achieve a 2-6\% SE improvement. Again, Border's SE and fairness are lower than HybridUA while requiring more fronthaul signaling.
	
	We also evaluate the impact of area size in the third and last groups of bars in Fig. \ref{fig:varying-CPU-SE}, where we double the area from 8 to 16$\text{km}^2$ to make the density of APs and UEs more sparse. When the area size increases, HybridUA suffers the second least performance degradation (11\%, after LLSFB at 8\%). As a result, HybridUA's SE is now on par with other methods while still consuming the least fronthaul signaling. SCF2 and Border suffer the most with 22\% SE loss.
	
	\subsection{Discussion}\label{sec:sim_discussion} 
	We found that a only very small number of APs make a strong contribution to the service of a UE (Fig. \ref{fig:varying-K-AP-per-UE}). This makes our hybrid network- and user-centric clustering approach a viable option that is particularly attractive to systems with limited fronthaul capacity. The other benchmark approaches have not taken advantage of this observation. The network-centric schemes (Nearest and LLSFB) are not terribly bad and may even be acceptable for extreme cases with little to no inter-CPU fronthaul capacity; but their SE, fairness, and 5\% outage (Fig. \ref{fig:varying-K-CDF}) can be significantly improved by partially incorporating user-centric clustering like we do. 
	
	In terms of scalability, our approach meets all the criteria of a scalable CFmMIMO system as defined in \cite{bjornson2020scalable}. It is even much more scalable than SCF \cite{bjornson2020scalable} and SCF1lim \cite{freitas2023reducing} measured by the processing power required on CPUs (Fig. \ref{fig:varying-K-UE-per-CPU}), which is quantified by the number of UEs that a CPU has to serve. Our approach is also highly flexible in handling the trade-off between SE and fronthaul load through adjusting the z-score threshold $ \epsilon $ - the higher the $ \epsilon $, the fewer "cell-edge" UEs, the lower the SE and the greater the saving of fronthaul signaling, and vice versa.
	
	The robustness of our approach is also highlighted in Fig. \ref{fig:varying-CPU-SE} when we increase the geographical area size. While other approaches experience a big SE loss, our approach still fares very well and becomes the best approach. 
	
	\section{Conclusion and Future Work}\label{sec:conclusion}
	We proposed a hybrid network- and user-centric online user association scheme for CFmMIMO that is highly scalable and robust. We showed that it is a viable option worth considering as we can save a considerable amount of inter-CPU fronthaul signaling and CPU processing power with a small loss, or no loss in some cases, of spectral efficiency and fairness.
	
	\textcolor{\changehighlightone}{Following this work, there are several potential research directions. With the distributed implementation of CFmMIMO, there might exist a correlation between the load per AP to its CPU and the fronthaul loads between the CPUs. A theoretical quantification of this correlation could be an interesting contribution. Some other avenues of interest for CFmMIMO include the integration of machine learning techniques for adaptive clustering or the impact of mobility on UE-AP association in dynamic environments.}
	
	\bibliography{IEEEabrv,bibliography}

\begin{thebibliography}{10}
\providecommand{\url}[1]{#1}
\csname url@samestyle\endcsname
\providecommand{\newblock}{\relax}
\providecommand{\bibinfo}[2]{#2}
\providecommand{\BIBentrySTDinterwordspacing}{\spaceskip=0pt\relax}
\providecommand{\BIBentryALTinterwordstretchfactor}{4}
\providecommand{\BIBentryALTinterwordspacing}{\spaceskip=\fontdimen2\font plus
\BIBentryALTinterwordstretchfactor\fontdimen3\font minus
  \fontdimen4\font\relax}
\providecommand{\BIBforeignlanguage}[2]{{%
\expandafter\ifx\csname l@#1\endcsname\relax
\typeout{** WARNING: IEEEtran.bst: No hyphenation pattern has been}%
\typeout{** loaded for the language `#1'. Using the pattern for}%
\typeout{** the default language instead.}%
\else
\language=\csname l@#1\endcsname
\fi
#2}}
\providecommand{\BIBdecl}{\relax}
\BIBdecl

\bibitem{bjornson2020scalable}
E.~Bj{\"o}rnson and L.~Sanguinetti, ``{Scalable cell-free massive MIMO
  systems},'' \emph{{IEEE} Trans. Commun.}, vol.~68, no.~7, pp. 4247--4261,
  2020.

\bibitem{ngo2024ultra}
H.~Q. Ngo, G.~Interdonato, E.~G. Larsson, G.~Caire, and J.~G. Andrews,
  ``{Ultra-dense cell-free massive MIMO for 6G: Technical overview and open
  questions},'' \emph{arXiv preprint arXiv:2401.03898}, 2024.

\bibitem{ngo2017cell}
H.~Q. Ngo, A.~Ashikhmin, H.~Yang, E.~G. Larsson, and T.~L. Marzetta,
  ``{Cell-free massive MIMO versus small cells},'' \emph{{IEEE} Trans. Wireless
  Commun.}, vol.~16, no.~3, pp. 1834--1850, 2017.

\bibitem{buzzi2017cell}
S.~Buzzi and C.~D’Andrea, ``{Cell-free massive MIMO: User-centric
  approach},'' \emph{{IEEE} Wireless Commun. Lett.}, vol.~6, no.~6, pp.
  706--709, 2017.

\bibitem{demir2021foundations}
{\"O}.~T. Demir, E.~Bj{\"o}rnson, L.~Sanguinetti \emph{et~al.}, ``{Foundations
  of user-centric cell-free massive MIMO},'' \emph{Foundations and Trends in
  Signal Processing}, vol.~14, no. 3-4, pp. 162--472, 2021.

\bibitem{guenach2020joint}
M.~Guenach, A.~A. Gorji, and A.~Bourdoux, ``{Joint power control and access
  point scheduling in fronthaul-constrained uplink cell-free massive MIMO
  systems},'' \emph{{IEEE} Trans. Commun.}, vol.~69, no.~4, pp. 2709--2722,
  2020.

\bibitem{zaher2022learning}
M.~Zaher, {\"O}.~T. Demir, E.~Bj{\"o}rnson, and M.~Petrova, ``{Learning-based
  downlink power allocation in cell-free massive MIMO systems},'' \emph{{IEEE}
  Trans. Wireless Commun.}, vol.~22, no.~1, pp. 174--188, 2022.

\bibitem{kassam2023joint}
J.~Kassam, D.~Castanheira, A.~Silva, R.~Dinis, and A.~Gameiro, ``{Joint
  Decoding and UE-APs Association for Scalable Cell-Free Systems},''
  \emph{{IEEE} Trans. Commun.}, 2023.

\bibitem{ammar2021downlink}
H.~A. Ammar, R.~Adve, S.~Shahbazpanahi, G.~Boudreau, and K.~V. Srinivas,
  ``{Downlink resource allocation in multiuser cell-free MIMO networks with
  user-centric clustering},'' \emph{{IEEE} Trans. Wireless Commun.}, vol.~21,
  no.~3, pp. 1482--1497, 2021.

\bibitem{interdonato2019scalability}
G.~Interdonato, P.~Frenger, and E.~G. Larsson, ``{Scalability aspects of
  cell-free massive MIMO},'' in \emph{IEEE International Conference on
  Communications (ICC)}.\hskip 1em plus 0.5em minus 0.4em\relax IEEE, 2019, pp.
  1--6.

\bibitem{ranjbar2022cell}
V.~Ranjbar, A.~Girycki, M.~A. Rahman, S.~Pollin, M.~Moonen, and E.~Vinogradov,
  ``{Cell-free mMIMO support in the O-RAN architecture: A PHY layer perspective
  for 5G and beyond networks},'' \emph{IEEE Communications Standards Magazine},
  vol.~6, no.~1, pp. 28--34, 2022.

\bibitem{freitas2023reducing}
M.~M. Freitas, D.~D. Souza, D.~B. da~Costa, A.~M. Cavalcante, L.~Valcarenghi,
  G.~S. Borges, R.~Rodrigues, and J.~C. Costa, ``{Reducing Inter-CPU
  Coordination in User-Centric Distributed Massive MIMO Networks},''
  \emph{{IEEE} Wireless Commun. Lett.}, 2023.

\bibitem{li2023joint}
Z.~Li, F.~G{\"o}ttsch, S.~Li, M.~Chen, and G.~Caire, ``{Joint Fronthaul Load
  Balancing and Computation Resource Allocation in Cell-Free User-Centric
  Massive MIMO Networks},'' \emph{arXiv preprint arXiv:2310.14911}, 2023.

\bibitem{wei2022user}
C.~Wei, K.~Xu, X.~Xia, Q.~Su, M.~Shen, W.~Xie, and C.~Li, ``{User-centric
  access point selection in cell-free massive MIMO systems: A game-theoretic
  approach},'' \emph{{IEEE} Commun. Lett.}, vol.~26, no.~9, pp. 2225--2229,
  2022.

\bibitem{guo2021joint}
F.~Guo, H.~Lu, and Z.~Gu, ``{Joint power and user grouping optimization in
  cell-free massive MIMO systems},'' \emph{{IEEE} Trans. Wireless Commun.},
  vol.~21, no.~2, pp. 991--1006, 2021.

\bibitem{vu2020joint}
T.~X. Vu, S.~Chatzinotas, S.~ShahbazPanahi, and B.~Ottersten, ``{Joint power
  allocation and access point selection for cell-free massive MIMO},'' in
  \emph{IEEE International Conference on Communications (ICC)}.\hskip 1em plus
  0.5em minus 0.4em\relax IEEE, 2020, pp. 1--6.

\bibitem{zhang2024rate}
Y.~Zhang, H.~Zhao, Y.~Mao, W.~Xia, W.~Lu, and H.~Zhu, ``{Rate-Splitting
  Multiple Access in Cell-Free Massive MIMO-URLLC Systems: Achievable Rate
  Analysis and Optimization},'' \emph{{IEEE} Trans. Commun.}, 2024.

\bibitem{zhang2024performance}
Y.~Zhang, W.~Xia, H.~Zhao, Y.~Zhu, W.~Xu, and W.~Lu, ``{Performance analysis of
  cell-free massive MIMO-URLLC systems over correlated Rician fading channels
  with phase shifts},'' \emph{{IEEE} Trans. Wireless Commun.}, 2024.

\bibitem{bjornson2017massive}
E.~Bj{\"o}rnson, J.~Hoydis, L.~Sanguinetti \emph{et~al.}, ``{Massive MIMO
  networks: Spectral, energy, and hardware efficiency},'' \emph{Foundations and
  Trends{\textregistered} in Signal Processing}, vol.~11, no. 3-4, pp.
  154--655, 2017.

\bibitem{ngo2017total}
H.~Q. Ngo, L.-N. Tran, T.~Q. Duong, M.~Matthaiou, and E.~G. Larsson, ``{On the
  total energy efficiency of cell-free massive MIMO},'' \emph{IEEE Transactions
  on Green Communications and Networking}, vol.~2, no.~1, pp. 25--39, 2017.

\bibitem{bjornson2011optimality}
E.~Bjornson, N.~Jalden, M.~Bengtsson, and B.~Ottersten, ``{Optimality
  properties, distributed strategies, and measurement-based evaluation of
  coordinated multicell OFDMA transmission},'' \emph{{IEEE} Trans. Signal
  Process.}, vol.~59, no.~12, pp. 6086--6101, 2011.

\end{thebibliography}
\end{document}